# Magnetic Control of Transmission and Helicity of Nano-Structured Optical Beams in Magnetoplasmonic Vortex Lenses

Nicolò Maccaferri,[1,a)] Yuri Gorodetski,[2,b)] Andrea Toma[1], Pierfrancesco Zilio[1], Francesco De Angelis[1], and Denis Garoli[1,c)]

[1]*Istituto Italiano di Tecnologia, Via Morego 30, 16136 Genova, Italy*

[2]*Mechanical Engineering Department and Electrical Engineering Department, Ariel University, 40700, Ariel, Israel*

[a)]nicolo.maccaferri@iit.it  [b)]yurig@ariel.ac.il  [c)]denis.garoli@iit.it

**We theoretically investigate the generation of far-field propagating optical beams with a desired orbital angular momentum by using an archetypical magnetoplasmonic tip surrounded by a gold spiral slit. The use of a magnetic material can lead to important implications once magneto-optical activity is activated through the application of an external magnetic field. The physical model and the numerical study presented here introduce the concept of magnetically tunable plasmonic vortex lens, namely a magnetoplasmonic vortex lens, which ensures a tunable selectivity in the polarization state of the generated nanostructured beam. The presented system provides a promising platform for a localized excitation of plasmonic vortices followed by their beaming in the far-field with an active modulation of both light's transmittance and helicity.**

In the recent years, structured optical beams have become a subject of an intense research[1], due to numerous potential applications they offer in the fields of super-resolution imaging[2], optical tweezing[3], nano-manipulation[4] and telecommunications[5]. A special interest is dedicated to the interaction of structured light with metallic nanostructures, resulting in Surface Plasmon Polaritons (SPPs) carrying optical angular momentum (OAM)[6-13]. These surface confined electromagnetic distributions are generally defined by a field singularity surrounded by a helical phase front, referred here as plasmonic vortex (PV). The singularity strength is named the topological charge of the vortex. The latter is defined by the phase ramp acquired in one round trip around the singularity center. This charge is proportional to the OAM carried by the field[8]. PVs are usually generated by coupling propagating beams to the plasmonic mode on a metal surface by using centro-symmetric coupling structure, known also as plasmonic vortex lenses (PVLs)[8]. A PVL generally comprises of periodic spiral or circular grooves or slits milled in the metal[6-17]. The total OAM of the resulting PV carries contributions from both the incident OAM and the additional OAM introduced by the coupling structure. More complex architectures proposed so far allow both coupling of light with a PV and out-coupling of the propagating PV into free-space, generating an out-coming nanostructured beam carrying non-zero OAM[7,13,15,16,18,19]. In this case, the PV generated by interaction of the incident light with a PVL can be modulated in the near-field and then scattered into free space by a proper decoupling structure. Most of the proposed systems utilized scattering as a primary coupling/decoupling mechanism[7,16,18]. Although it was shown that this method provides a way to abruptly modify the out-coming OAM[18], it demonstrates a very low transmission efficiency and non-pure polarization state of the transmitted light. Recently, an efficient PV coupling to the free space was experimentally demonstrated by means of a single-layer PVL structure via adiabatically tapered gold tip at its center[13]. In Ref. 13 it was shown that by properly shaping the nanotip geometry the PV excited at the spiral structure can be coupled to far-field distribution carrying well defined OAM with a transmission efficiency of up to 90%. Moreover, this type of system ensures a polarization selectivity of about 90% independently on the original polarization state of the incident light[20]. For instance, if the incident light is either right hand or left hand circularly polarized (RCP or LCP, respectively), the out-coming beam shows an almost pure RCP state[20]. Nevertheless, these intriguing effects can be solely controlled by the properly designed 3D shape of the tip and by the illuminating beam parameters, due to the passive nature of the metal used – gold in this specific case. Various practical applications in nanophotonics require fast external control of the emerging beam characteristics. In this regard magnetoplasmonic devices draw a very promising route to active nanophotonics, since an externally applied magnetic field can alter the plasmonic response through the activation of magneto-optical (MO) effects inside the material, leading to novel and unexpected effects[21-25]. The MO effect is a physical phenomenon taking place inside a magnetic material that can break locally the symmetry of time inversion leading to a modification of the polarization and the intensity of the light interacting with such magneto-optical materials, and has been widely used in industry over the last 30 years. The most widely applications are magneto-optical recording[26,27] and magneto-optical isolators[28,29]. Obviously, the main requirement of a practical magnetoplasmonic device is to be fabricated from a material that has both good plasmonic and magnetic properties, for instance hybrid noble metal/magnetic materials[30-32]. In this Letter, we theoretically present and analyze a novel architecture where a PV excited by an archetypical PVL milled in a gold surface propagates on an adiabatically tapered tip made of a composite magnetically active material and detaches to the far-field while carrying a well-defined OAM. While a tapered plasmonic nanotip was recently demonstrated to produce a very efficient beaming of vortex modes with clearly defined orbital and spin angular momentum states by adding a magnetic tunability we propose to convert this device to active and externally controllable. We analyze the out-coming light transmission efficiency and helicity, showing that, despite generally higher losses of magnetoplasmonic tip material with respect to a noble metal-based tip, our 3D structure exhibits relatively high energy throughput. Moreover, we show that the transmission efficiency and helicity locking effect of our magnetoplasmonic PVL (MPVL) can be modulated by applying an external magnetic field, providing optimistic opportunities for the realization and the exploitation of magnetically-driven PVs. In order to perform an optimal and, at the same time, fair study, we start analyzing the beam characteristics of a PVL similar to that studied in Ref. 13 and Ref. 20 at $\lambda$ = 780nm. It is worth mentioning that at this wavelength, and more in general in the near-IR spectral region, the MO response of the magnetic material chosen for our calculations, viz. a CoFeTb alloy, is large enough to induce noticeable changes in the analyzed light characteristics due to its inherent strong spin-orbit coupling[33,34]. In Fig. 1(a) we depict the scheme of the proposed structure, which comprises a multiple-turn spiral slit milled on a 300 nm Au film deposited onto a 100 nm $Si_3N_4$ membrane and a conical metallic tip made of a magnetoplasmonic material located exactly at the center of the spiral. It is worth mentioning here that the fact that a high-index substrate $Si_3N_4$ is used might make this lens very promising for integration with conventional photonic integrated circuits that can be manufactured at the operational wavelength in a waveguide based system. Moreover, from a fabrication point of view, the protocol to fabricate this kind of devices is already well established in the community[13]. The deposition of more than one material can be done, using the well-established DC-Magnetron Sputtering Deposition Technique, which allows for a fine tuning of the CoFeTb composition.

In our calculations, we consider an effective average medium that combines the dielectric tensor of Au and that of CoFeTb, giving rise to non-diagonal dielectric tensor once a magneto-optical activity is activated within the tip. The nonzero off-diagonal elements depend on the relative orientation of the geometry, of the exciting radiation, and of the magnetic field. In our case, the external magnetic field is aligned along the tip (i.e., aligned along the z-direction; see Fig. 1(a)), and the dielectric tensor presents the form

$$\begin{pmatrix} \varepsilon_d & \varepsilon_{off} & 0 \\ -\varepsilon_{off} & \varepsilon_d & 0 \\ 0 & 0 & \varepsilon_d \end{pmatrix} \tag{1}$$

Depending on the amount of magnetic material within the tip, the elements of the dielectric tensor read as

$$\varepsilon_d = f\varepsilon_{d,Au} + (1-f)\varepsilon_{d,CoFeTb}; \ \varepsilon_{off} = (1-f)\varepsilon_M \tag{2}$$

Here $f$ is the relative amount of Au within the tip and $\varepsilon_M = i\varepsilon_{d,CoFeTb}K$, where $K$ is the so-called magneto-optical Voigt constant accounting for spin-orbit coupling within the magnetic material. Moreover, three main parameters define the tip shape, namely its height (h = 6500nm), aperture angle (α = 20°), curvature radius at the basis ($r_c$ = 1470nm) and the curvature radius at the tip apex (50 nm). These parameters are kept fixed throughout the manuscript.

First, we analyze the case of a circularly polarized plane-wave impinging normally from the Si3N4 side, considering a zero magnetization within the tip, namely $\varepsilon_{off} = 0$. In this case the dielectric constant of the magnetoplasmonic tip is given by the single complex scalar $\varepsilon_d$ defined in Eq. (2). As described elsewhere, the electric field component locally orthogonal to the slits efficiently couples to the SPP mode of the metal-air interface[6,8,14]. The plasmonic wave fronts launched by each spiral constructively interfere producing a PV radially propagating towards the PVL center. In the absence of the tip, the PV confined by the spiral grooves forms a standing wave, giving rise to a characteristic Bessel interference pattern. The z-component of the electric field is then given in cylindrical coordinates by[8]

$$E_{z,l}(r,\varphi,z) = AJ_l(k_{SPP}r)e^{-kz}e^{il\varphi} \tag{3}$$

where $k_{SPP}$ is the wave vector of an SPP propagating on a flat gold-air interface, $J_l$ is the $l^{th}$ order Bessel function of first kind and $k = \sqrt{k_{SPP}^2 - k_0^2}$, where $k_0 = 2\pi/\lambda_0$ is the wave vector of light in vacuum.

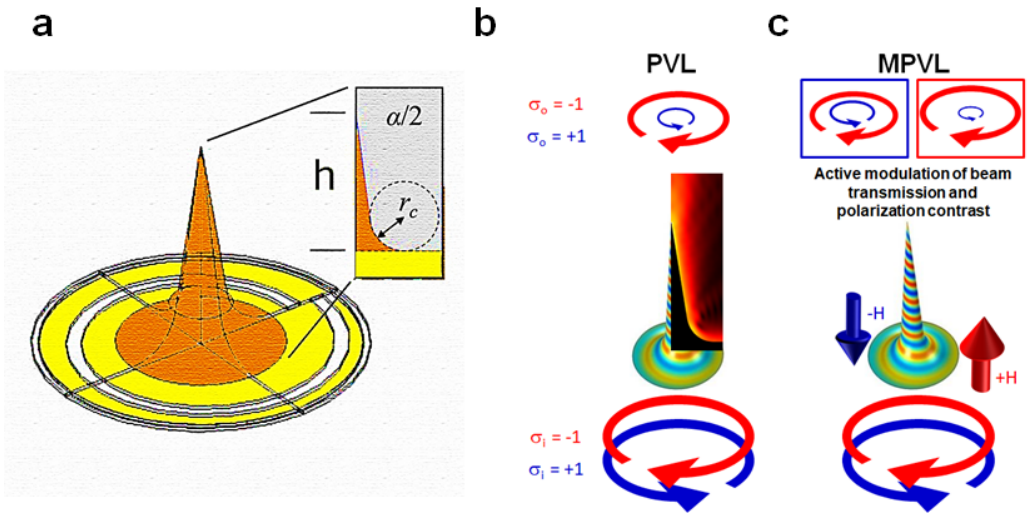

**Fig. 1.** (a) Sketch of the architecture studied in this work: a spiral slit on top of a Si3N4 membrane with at its center a magnetoplasmonic tip. (b) Illustration of the helicity locking concept induced by a PVL based on an adiabatically tapered plasmonic tip. (c) Pictorial representation of the magnetic modulation of transmission and helicity locking via a MPVL.

It can be shown that the PV carries an OAM proportional to the PV topological charge $l$. The latter is given by $l = m + \sigma_i$ with the incident light spin number $\sigma_i = -1$ and $\sigma_i = +1$ corresponding to the RCP and LCP light, respectively, and $m$ defining the topological charge of the spiral[13]. When the conical tip is present at the PVL center the surface confined electromagnetic mode may couple to the guided mode propagating along the tip upwards. In Fig. 1(b) we show the $E_z$ component of the plasmonic field (simulated using the Finite Element Method implemented in COMSOL Multiphysics Software) for the case $l = 1$ in an Au tip. In this case, most of the energy is smoothly guided towards the top of the tip, and the spiral phase-fronts of the PV propagate on the tip almost without losses. This guided mode experiences gradual acceleration due to a local increasing of the effective index, and at some height, where its momentum matches the one of the free-space radiation, it detaches to the far-field[13]. Interestingly, as already demonstrated in Ref. 20, the emerging polarization handedness $\sigma_o$ is almost purely RCP. Thus, a PVL made of pure Au presents a strong chiral nature and allows the helicity locking of the out-coming beam to a particular polarization state.

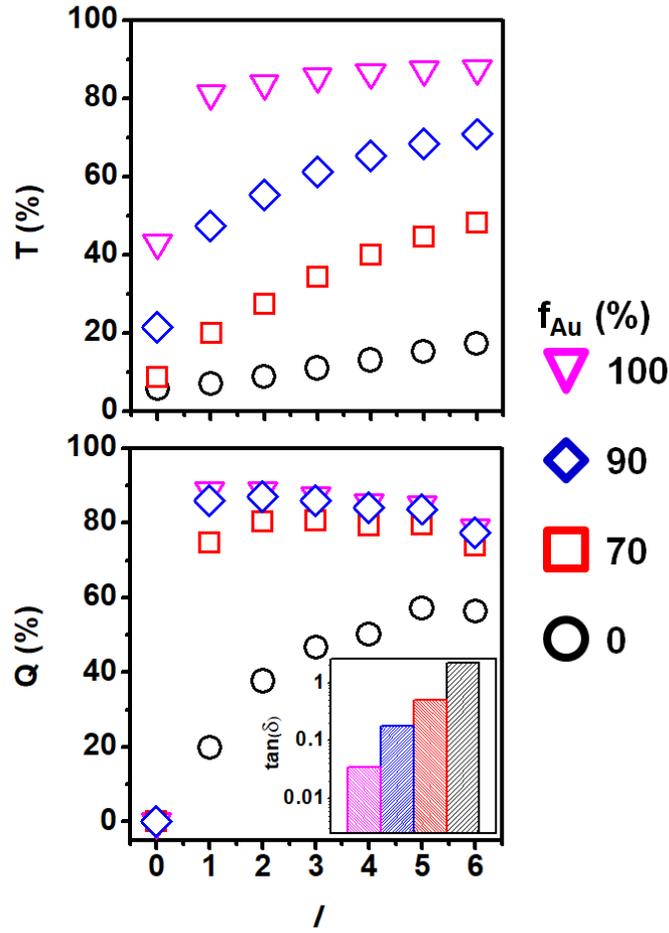

**Figure 2.** Transmittance T (top-panel) and polarization contrast Q (bottom-panel) as function of the topological charge $l$ of the PV. Inset: dielectric loss $\tan(\delta)$.

The main focus of this Letter is to demonstrate that this effect, as well as the transmittance T of the out-coming beam, can be actively modulated using an external magnetic field when the tip is made of a magnetoplasmonic material. It is well known that, if a magnetic field H is applied along the propagation direction of a chiral wave front the orientation of H with respect to the wave vector of the wave front can favor or not one particular light spin number (Fig. 1(c))[35]. This effect allows an active modulation of the transmittance of the tip, as well as of the so-called polarization contrast. The latter is defined as $Q = 1 - P_+/P_-$, with $P_+$ and $P_-$ being the light powers decoupled by the tip with RCP and LCP state, respectively. It is worth noticing here that relatively small (<0.5 T) magnetic fields are required to saturate the magnetization inside the CoFeTb alloy. A smooth connection of the tip base ensures almost lossless coupling of the plasmonic mode at the surface to the mode guided at the cone. The only dielectric losses in the system are attributed to the plasmonic absorption in the metal. By replacing gold with a magnetic material we introduce higher ohmic losses. As shown in the bottom-panel of Fig. 2, the dielectric loss, defined as $\tan(\delta) = Im(\varepsilon_d)/Re(\varepsilon_d)$[36], of pure CoFeTb alloy bulk material is > 1, leading to a very low transmission efficiency (T < 20%) and polarization contrast ($Q$ < 60% for $l > 1$), as shown in the top and bottom-panel of Fig. 2, respectively, at $\lambda_0$ = 780 nm. The same effects achieved by using a gold tip (T and Q > 80% for $l > 1$) can be partially retrieved using a combination of Au and CoFeTb. We consider two cases, namely $f_{Au}$ = 90% and $f_{Au}$ = 70%. In both cases the losses are reduced by almost one order of magnitude ($\tan(\delta)$ < 1).

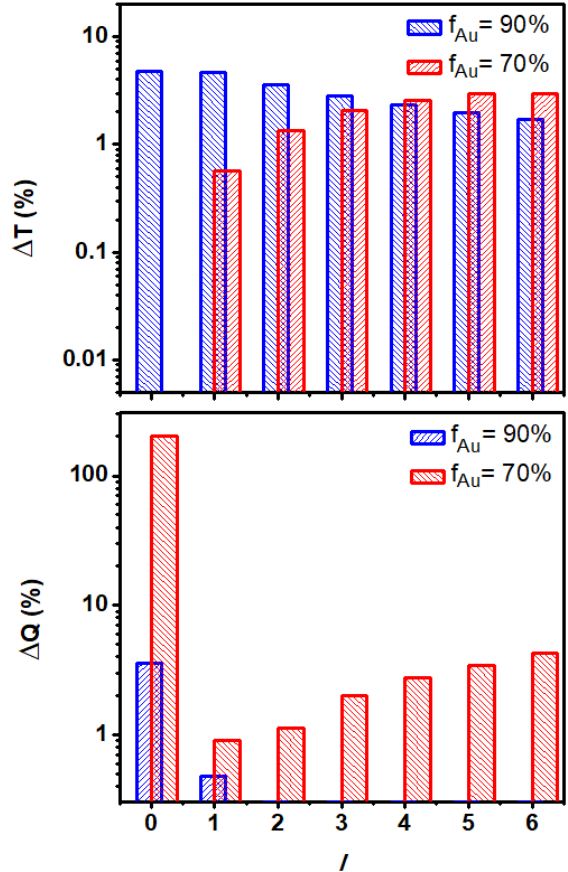

**Figure 3.** Magnetic modulation of T (top-panel) and Q (bottom-panel) for $f_{Au}$ = 90% (red bars) and $f_{Au}$ = 70% (blue bars); y-scale is $\log_{10}$.

This strategy leads to a straightforward improvement of the transmission and polarization contrast efficiencies, and we start approaching the results obtained for a pure Au tip. Interestingly, while T remains below 70% also for $f_{Au}$ = 90%, $Q$ is almost the same of the pure Au case for both $f_{Au} = 90\%$ and $f_{Au} = 70\%$. We have so demonstrated that a MPVL made of Au and CoFeTb allows not only a relatively high energy throughput, but also preserve the outstanding property of locking one particular helicity state of the out-coming beam (in this case an almost complete RCP state).

Finally, we come at the more interesting part of this work, where we activate the magneto-optical properties of the tip by considering a nondiagonal dielectric tensor, as reported in Eq. (1). Here we assume that the magnetization within the tip is in the saturated state. The modulation of both T and Q, namely ΔT and ΔQ, can be defined as the overall variation of T and Q between the two magnetization saturation states at –H and at +H, namely ΔT = 1 – [T(+H)]/[T(–H)] and ΔQ = 1 – [Q(+H)]/[Q(–H)]. As it can be inferred by the top-panel of Fig. 3, the modulation of the transmission is larger than 1% (except for the case $l = 0$, where nevertheless there is no coupling of the PV to the far-field) for both the concentrations of Au considered here. It is worth noticing that this modulation is pretty high considering that it is achieved using a quite low concentration (<50%) of active magnetic material within the tip. Also in the case of ΔQ, where the higher modulations are reasonably obtained for the case where $f_{Au} = 70\%$, we exceed the 1%. It is important to mention that using common transition metals as magnetic material, such as Fe, Co or Ni, the modulation effects presented here are reduced by at least one order of magnitude. Moreover, we predict that this modulation effect can be achieved by using relative small values of the external magnetic field, for sure not exceeding 0.5 T. Finally, it is worth noticing that for $l = 0$ a modulation of almost 200% is obtained, since in this case Q at H=0 is zero. Although in this latter case we obtain a very large value for ΔQ, the effect is rather small (we go from Q=0 at H=0 to Q=$10^{-3}$ at H≠0). Nevertheless this interesting case shows the huge potential of having magneto-optically active materials within plasmonic or photonic devices to induce physical effects which are not present in a device made of a passive material, such as gold. In this context the currently presented ability of an external optical state control could be a key feature in such nanophotonic applications as nanoscale local polarization detection, chirality recognition and polarization spectroscopy, as well as magnetic field sensing or tunable near-field emission of a desired optical state. Moreover, we believe that our proposed scheme can lay in the basis of more general effect – magnetoplasmonic topology control – which can be integrated for quantum computing, sensing, optical communication etc.

In summary, we have studied a plasmonic vortex lens structure capable of coupling a circularly polarized light to a plasmonic vortex and efficiently transmitting it to the far-field by means of an adiabatically tapered magnetoplasmonic tip placed at its center. The dimensions required are feasible for nano-fabrication and integration in real world devices. The higher losses related to the use of a magnetoplasmonic material instead of a pure noble metal material lead to lower transmittance to the far-field, although a high polarization contrast (> 80%) is retained also in the case where the tip is made of a magnetoplasmonic material. Finally, we have theoretically demonstrated that a magnetic modulation of both transmittance and polarization contrast of more than 1% is possible by applying an external magnetic field with a relative small amount of magnetic material within the tip (<50%).

## ACKNOWLEDGEMENTS

N. M., P. Z., F. D. A. and D. G. acknowledge support from the European Research Council under the European Union's Seventh Framework Programme (FP/2007-2013) – ERC Grant Agreement no. [616213], CoG: Neuro-Plasmonics, and under the Horizon 2020 Program, FET-Open: PROSEQO, Grant Agreement no. [687089]. Y. G. acknowledges support from the Ministry of Science Technology & Space, Israel. A.T is grateful for financial support from Compagnia di San Paolo under grant agreement ID ROL 10262.